\documentclass[aps,prc,twoside,twocolumn,nofootinbib,
showpacs,floatfix,dvips]{revtex4}

\usepackage{amsmath,amssymb,graphicx,bm}

\begin{document}


\title{\boldmath
Nuclear modification factor of non-photonic electrons in heavy-ion
collisions and the heavy-flavor baryon to meson ratio}


\author{Yongseok Oh}%
\email{yoh@comp.tamu.edu}

\affiliation{Cyclotron Institute and Physics Department,
Texas A\&M University, College Station, Texas 77843, U.S.A.}

\author{Che Ming Ko}%
\email{ko@comp.tamu.edu}

\affiliation{Cyclotron Institute and Physics Department,
Texas A\&M University, College Station, Texas 77843, U.S.A.}

\date{\today}


\begin{abstract}

The nuclear modification factor $R_{AA}$ of non-photonic electrons
in $\mbox{Au} + \mbox{Au}$ collisions at $\sqrt{s_{NN}^{}} =
200$~GeV is studied by considering the decays of heavy-flavor
hadrons produced in a quark coalescence model. Although an
enhanced $\Lambda_c/D^0$ ratio is predicted by the coalescence
model, it is peaked at small transverse momenta ($\sim 2$~GeV) due
to the large difference between heavy and light quark masses. As a
result, the enhanced $\Lambda_c/D^0$ ratio, which is expected
to suppress the electron $R_{AA}$ as the branching ratio of
$\Lambda_c$ decay into electrons is smaller than that of $D^0$,
does not lead to additional suppression of the electron $R_{AA}$ at
large transverse momenta ($\ge 5$~GeV), where the suppression is
mainly due to heavy quark energy loss in produced quark-gluon
plasma. Also, the enhanced $\Lambda_b/\bar{B}^0$ ratio predicted by
the coalescence model has even smaller effect on the non-photonic
electron $R_{AA}$ as bottom baryons and mesons have similar
branching ratios for semi-leptonic decays into electrons.

\end{abstract}


\maketitle


Heavy-flavor hadrons are useful for probing the properties of the
dense matter produced in relativistic heavy-ion collisions. Although
$D$ mesons were measured in $d + \mbox{Au}$ collisions at the
Relativistic Heavy Ion Collider (RHIC)~\cite{STAR04d}, there is not
yet direct measurement of open charm or open bottom hadrons in
heavy-ion collisions. Instead, heavy-flavor hadron production in
heavy ion collisions was studied through measurements of
non-photonic electrons from their decays~\cite{PHENIX05f}. The
transverse momentum ($p_T^{}$) spectrum of these electrons in
$\mbox{Au} + \mbox{Au}$ collisions at $\sqrt{s_{NN}^{}}=200$~GeV
were measured for several centralities by the PHENIX
Collaboration~\cite{PHENIX06}. The resulting nuclear modification
factor $R_{AA}$, defined by the ratio of the $p_T^{}$ spectrum of
non-photonic electrons in heavy ion collisions $dN_{AA}^e/dp_T^{}$
to that from proton-proton collisions $dN_{pp}^e/dp_T^{}$ multiplied
by the initial number of $NN$ binary collisions $\langle N_{\rm
coll}^{AA} \rangle$, i.e.,
\begin{equation} R_{AA} =
\frac{dN^e_{AA}/dp_T^{}}{\langle N_{\rm coll}^{AA} \rangle \,
dN_{pp}^e/dp_T^{}}, \label{eq:def_RAA}
\end{equation}
shows that the production of heavy-flavor hadrons
in heavy-ion collisions is suppressed as much as
that of pions. This has raised a challenging question on heavy
quark interactions in medium as perturbative Quantum Chromodynamics
(pQCD) predicts that heavy quarks lose less energy in the
quark-gluon plasma than gluons and light quarks, which is
responsible for the observed large suppression of high $p_T^{}$
pions. Various ideas have been suggested to explain the observed
small non-photonic electron $R_{AA}$, and these include the
introduction of very large medium opacity~\cite{ACDSW05},
two-body~\cite{Mustafa05,VGR06,ZCK05} and three-body collisional
energy loss~\cite{LK07}, and the collisional dissociation of heavy
mesons in medium~\cite{AV06}. It is also proposed that an
enhancement of the $\Lambda_c/D^0$ ratio would reduce the $R_{AA}$
of non-photonic electrons~\cite{SD05,MGC07}. This is based on the
observation that the branching ratio for $\Lambda_c \to e$ decay is
smaller than that for $D^0 \to e$ decay. Therefore, if the
$\Lambda_c/D^0$ ratio is enhanced as in the case of $p/\pi$ and
$\Lambda/K^0$ ratios, the $R_{AA}$ of non-photonic electrons would
be smaller than that only due to heavy quark energy loss. In
Ref.~\cite{MGC07}, it was claimed that this effect may explain about
20\% of the suppression of $R_{AA}$. Since the origin of the
$\Lambda_c/D^0$ enhancement is unknown, this ratio is assumed in
Refs.~\cite{SD05,MGC07} to have either the same $p_T^{}$ dependence
as in the $\Lambda/K$ data or a Gaussian form.

Recently, the enhancement of $\Lambda_c$ yield in relativistic heavy
ion collisions was suggested as a possible test of the diquark model
for the structure of heavy baryons~\cite{LOYYK07,OKLY09}. Estimates
based on the quark coalescence model for heavy-flavor hadron
production show that the existence of diquarks in both the
quark-gluon plasma and the $\Lambda_c$ would enhance the
$\Lambda_c/D^0$ ratio by a factor of ten and five, respectively,
compared to those of PYTHIA simulation for $pp$ collisions and of a
simple thermal model. (See, e.g., Ref.~\cite{GKR04} for the quark
coalescence model.) Furthermore, it was shown that the enhancement
of the $\Lambda_c/D^0$ ratio could be better seen at low $p_T^{}$
region where heavy quark fragmentation is less important. Therefore,
the enhancement of $\Lambda_c/D^0$ ratio is expected to result in
the suppression of $R_{AA}$ only at low $p_T^{}$ region. This is in
contrast to the assumption of Refs.~\cite{SD05,MGC07} that the
$\Lambda_c/D^0$ enhancement continues to the large $p_T^{}$ region.
In this report, we calculate the non-photonic electron $R_{AA}$
based on the results from Ref.~\cite{OKLY09}.  We also discuss the
contribution from the enhanced $\Lambda_b/\bar{B}^0$ ratio to the
non-photonic electron $R_{AA}$.

\begin{table*}[t]\centering
\begin{tabular}{c|cccc}
\hline\hline
Decay channel & $D^0$ & $D^+$ & $D_s^+$ & $\Lambda_c$ \\ \hline
$\mbox{BR}(e^+ + \mbox{anything})$ & $6.53 \pm 0.17\,\%$ & $16.0 \pm 0.4\,\%$ &
$ 8 \stackrel{+6}{-5}\,\%$ & $4.5 \pm 1.7\,\%$ \\ \hline
BR$(K e^+ \nu_e)$ & $3.58 \pm 0.06\,\%$ & $8.6 \pm 0.5\,\%$ & & \\
BR$(K^* e^+ \nu_e)$ & $2.18 \pm 0.16\,\%$ & $3.66 \pm 0.21\,\%$ & & \\
BR$(\eta \ell^+ \nu_\ell)$ & & & $2.9 \pm 0.6\,\%$ & \\
BR$(\eta' \ell^+ \nu_\ell)$ & & & $1.02 \pm 0.33\,\%$ & \\
BR$(\Lambda e^+ \nu_e)$ & & & & $2.1 \pm 0.6\,\%$ \\
BR$(pe^+ +\mbox{anything})$ & & & & $1.8 \pm 0.9\,\%$
\\ \hline
\end{tabular}
\caption{\label{tab:BR_c}%
Branching ratios of major semi-leptonic charm hadron decays into
electrons as listed by the Particle Data Group~\cite{PDG08}.}
\end{table*}

For calculating the electron spectrum from heavy hadron decays,
information on the branching ratios of semi-leptonic decays of heavy
hadrons into electrons is needed. The experimental data complied by
the Particle Data Group~\cite{PDG08} for the branching ratios of
these decays are, however, ambiguous as shown in
Table~\ref{tab:BR_c}. One can estimate the effect of the
$\Lambda_c/D^0$ enhancement on the electron $R_{AA}$ roughly by
considering semi-leptonic decay of heavy hadrons in each channel as
in Ref.~\cite{MGC07},
\begin{equation}\label{raa}
R_{AA} = \frac{\left\{ n_e(D^0) + n_e(D^+) + n_e(D_s) +
n_e(\Lambda_c)  \right\}_{AA}} {\left\{ n_e(D^0) + n_e(D^+) +
n_e(D_s) + n_e(\Lambda_c)  \right\}_{pp}},
\end{equation}
where $n_e(H)$ is the number of electron produced by the decay of
hadron $H$, and the number of binary collisions is included in the
numerator. Since the branching ratio of $\Lambda_c$ decay into
electron is smaller than that of charmed
mesons, the enhanced $\Lambda_c/D^0$ ratio indeed suppresses the
electron $R_{AA}$ if the momentum difference between electrons and
the decayed charmed hadrons is neglected. In Ref.~\cite{MGC07}, it
was claimed that if the enhancement factor, $C =
(\Lambda_c/D^0)_{AA}/(\Lambda_c/D^0)_{pp}$, is $12$, the electron
$R_{AA}$ is reduced by about 20\%. More quantitative understanding
of the effect of enhanced $\Lambda_c/D^0$ ratio on the $R_{AA}$ of
non-photonic electrons requires, however, a better understanding of
the branching ratios of heavy hadron decays, including the decays to
four-body final states, as well as the production mechanism of
heavy-flavor hadrons in both $pp$ and $AA$ collisions. The latter is
addressed in the present study.

\begin{figure}[t]\centering
\includegraphics[width=0.4\textwidth,angle=0,clip]{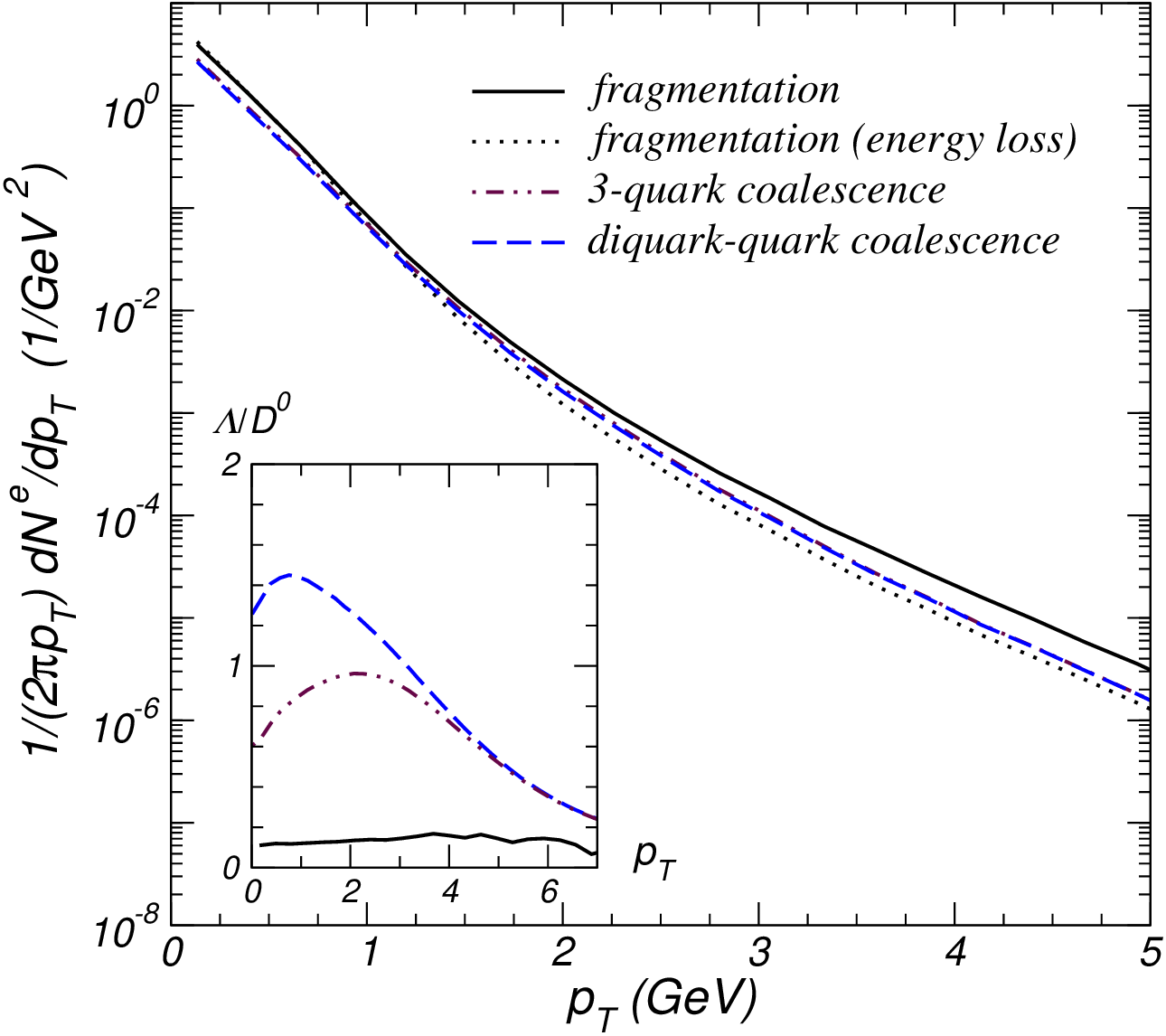}
\caption{\label{fig:charm-e} Spectra of electrons from decays of all
charmed hadrons. Shown in the inset are the $\Lambda_c/D^0$ ratios
in fragmentation and in both the three-quark and diquark-quark
coalescence models.}
\end{figure}

We use the coalescence model calculation presented in
Ref.~\cite{OKLY09}, which is based on the coalescence formulas
\begin{eqnarray}
\frac{dN_B}{d\bm{p}_{B}^{}} &=& g_{B}^{} \frac{(2\sqrt{\pi})^6
(\sigma_1^{} \sigma_2^{})^3}{V^2} \int d \bm{p}_1^{} d \bm{p}_2^{} d
\bm{p}_3 \frac{dN_1}{d\bm{p}_1^{}} \frac{dN_2}{d\bm{p}_2^{}}
\frac{dN_3}{d\bm{p}_3^{}} \nonumber \\ && \mbox{} \times \exp\left(
-\bm{k}_1^2 \sigma_1^2 - \bm{k}_2^2 \sigma_2^2 \right) \delta(
\bm{p}_{B}^{} - \bm{p}_1^{} - \bm{p}_2^{} - \bm{p}_3^{}),
\nonumber \\
\label{eq:3q-coal}
\end{eqnarray}
for baryon production from three-quark coalescence and
\begin{eqnarray}
\frac{dN_M}{d\bm{p}_M^{}} &=& g_M
\frac{(2\sqrt{\pi}\sigma_{dq}^{})^3}{V} \int d \bm{p}_1^{} d
\bm{p}_{2}^{} \frac{dN_1}{d\bm{p}_1^{}}
\frac{dN_{2}}{d\bm{p}_{2}^{}} \nonumber \\ && \mbox{} \times
\exp\left(-\bm{k}^2 \sigma_{dq}^2 \right) \delta( \bm{p}_B^{} -
\bm{p}_1^{} - \bm{p}_{2}^{}), \label{eq:dq-coal}
\end{eqnarray}
for both baryon and meson production from diquark-quark coalescence
and from quark--anti-quark coalescence, respectively. For the
details on the model and parameters used in the calculations, we
refer to Ref.~\cite{OKLY09}. For the heavy quark energy loss,
we use the parametrization given in Ref.~\cite{LK07},
\begin{eqnarray}
L_c &=& 0.8 \exp(-p/1.2) + 0.6 \exp(-p/15),
\nonumber \\
L_b &=& 0.36 + 0.84 \exp(-p/10),
\label{eq:loss}
\end{eqnarray}
for charm and bottom quarks, respectively, where $p$ is the
transverse momentum in units of GeV. As shown in Ref.~\cite{OKLY09},
compared with charmed hadron production via fragmentation of charm
quarks, the yield of $D^0$ mesons in heavy ion collisions is
suppressed in the quark coalescence model as a result of enhanced
production of $\Lambda_c$. (See the small window in
Fig.~\ref{fig:charm-e} for the $\Lambda_c/D^0$ ratios.) This happens
in models based on both the three-quark coalescence and the
diquark-quark coalescence. In Fig.~\ref{fig:charm-e}, we plot the
electron spectrum from charm hadron decays for central
$\mbox{Au}+\mbox{Au}$ collisions at mid-rapidity ($|y| \le 0.5$). As
a reference, results from the decays of charmed hadrons produced via
$\delta$-function fragmentation of charm quarks, i.e., charm quark
and produced hadrons have same $p_T^{}$, without and with energy
loss are given by the solid line and dashed line, respectively.
These electron spectra include contributions from decays of $D^0$,
$D^+$, $D_s$, $\Lambda_c$ and their anti-particles, as production of
electrons from charmed hadrons other than $D^0$ is not negligible,
particularly if there is a large enhancement of the
$\Lambda_c/D^0$ ratio~\cite{SD05}. Compared with
the case that these charmed hadrons are produced from fragmentation
of charm quarks via a $\delta$-function fragmentation, coalescence
models are seen to give a somewhat suppressed electron spectrum. We
note that the uncertainties in the branching ratios as shown in
Table~\ref{tab:BR_c} can affect the electron spectrum. More
discussions on these uncertainties can be found in
Ref.~\cite{Soren07}.

\begin{figure}[t]\centering
\includegraphics[width=0.4\textwidth,angle=0,clip]{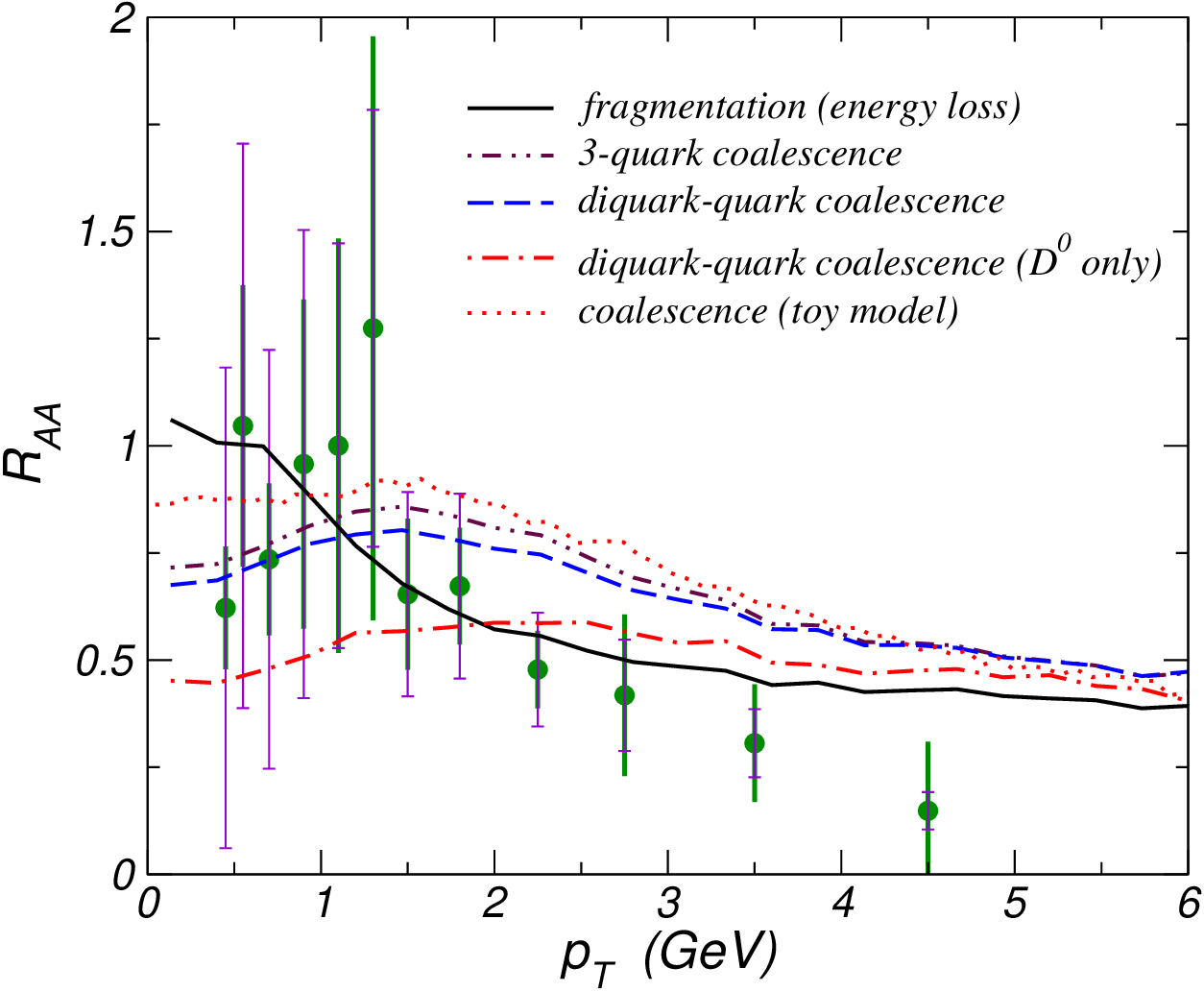}
\caption{\label{fig:RAA}%
The electron $R_{AA}$ in the central collision of Au$+$Au at
$\sqrt{s_{NN}^{}} = 200$~GeV from charmed hadrons. The solid line is
the result obtained from only fragmentation of heavy quarks with
energy loss. The dot-dot-dashed (dashed) line is for the three-quark
(diquark-quark) coalescence model, while the dot-dashed line is
obtained with $D^0$ meson only in the diquark-quark coalescence
model. The dotted line is from the toy model described in the text
and the data are from Ref.~\cite{PHENIX06}.}
\end{figure}

Our results for the non-photonic electron $R_{AA}$ are presented in
Fig.~\ref{fig:RAA}. Here, we consider five cases for investigating
the role of the enhancement of the $\Lambda_c/D^0$ ratio in the
electron $R_{AA}$. First, the solid line is the electron $R_{AA}$
for the case with charm quark energy loss in the $\delta$-function
fragmentation approximation as discussed in Ref.~\cite{LK07}. Shown
by the dotted line is the result from the coalescence model when
\emph{only} $D$ mesons are allowed to be produced. If other charmed
hadrons are allowed to be produced, the contribution from the $D^0$
mesons to the electron spectrum is suppressed and leads to the
suppressed electron $R_{AA}$ as can be seen by the dot-dashed line.
The dashed and dot-dot-dashed lines are obtained by allowing the
decays of all charmed hadrons into electrons in diquark and
three-quark coalescence, respectively, and are the main results of
this study. Therefore, by comparing the dotted line and the dashed
line, one can see that the enhanced $\Lambda_c/D^0$ leads to
suppression of the electron $R_{AA}$.

However, when we compare our result for the electron $R_{AA}$ with
that of fragmentation with energy loss, the enhancement of the
$\Lambda_c/D^0$ ratio does not cause the suppression of the electron
$R_{AA}$. This is because the enhancement of $\Lambda_c/D^0$ ratio in the coalescence model appears mainly in the
region of $p_T^{} \le 5$~GeV. In addition, we found that the
crossover of the coalescence and fragmentation is at $p_T^{} \sim
0.75$~GeV. As a result, the suppression of the $R_{AA}$ for
non-photonic electrons appears also at low $p_T^{}$ region ($p_T^{}
\le 3$~GeV). Thus, the calculated $R_{AA}$ underestimates the
data~\cite{PHENIX06} at low $p_T^{}$ and overestimates it at large
$p_T^{}$. Including the decay of $\Lambda_c$, whose branching ratio
of decaying into an electron is not small in comparison to that of
$D^0$, increases the electron $R_{AA}$. We also found that the
difference between the results of the three-quark and the
diquark-quark coalescence models is not large. Therefore, in the
coalescence model considered here, more quenched charm quark
spectrum is needed to explain the small electron $R_{AA}$ observed
in experiments.

Our results that the enhancement of $\Lambda_c/D^0$ ratio does not
cause the suppression of $R_{AA}$ is in sharp contrast with those of
Refs.~\cite{SD05,MGC07}. The reason for this is due to the
difference in the $\Lambda_c/D^0$ ratio at large $p_T^{}$ region. In
both Refs.~\cite{SD05,MGC07}, a constant enhancement factor is
assumed for the  $\Lambda_c/D^0$ ratio at large $p_T^{}$ region.
Besides the assumption that the $\Lambda_c/D^0$ ratio is the same as
the $\Lambda/K_S^0$ ratio, it is further assumed in Ref.~\cite{SD05}
that $\Lambda_c/D^0 = 0.33$ for large $p_T^{}$, where no data for
$\Lambda/K_S^0$ exist. In Ref.~\cite{MGC07}, an even larger constant
enhancement factor $C \simeq 12$ was introduced to the
$\Lambda_c/D^0$ ratio that was assumed to have a Gaussian shape in
$p_T^{}$. This large enhancement of $\Lambda_c/D^0$ at large
$p_T^{}$ would then lead to the suppression of $R_{AA}$ at
intermediate and large $p_T^{}$ ($\ge 5$~GeV).

In our case, although the $\Lambda_c/D^0$ ratio is enhanced compared
to that in $pp$ collisions, the enhancement is mainly at small
$p_T^{}$ and the $\Lambda_c/D^0$ ratio remains
comparable to that of $pp$ collisions at large $p_T^{}$ ($> 7 \sim
8$~GeV) region. Therefore, the effect of the $\Lambda_c/D^0$
enhancement on the electron $R_{AA}$ cannot be seen in the
intermediate electron $p_T^{}$ region. Furthermore, charmed hadrons
produced by coalescence with light quarks have larger values of
$p_T^{}$ than that of the charm quark inside a hadron, which results
in the shift of charmed hadron $p_T^{}$ spectra to the larger
$p_T^{}$ region. Therefore, at large $p_T^{}$ region, we have more
charm hadrons than those produced by $\delta$ function fragmentation
of charm quarks. Although this effect is not large compared to the
case for light hadrons, it increases the electron $R_{AA}$. As a
result, we have a larger $R_{AA}$ at intermediate electron $p_T^{}$
compared to that from the fragmentation of quenched heavy quarks.


As the electron $p_T^{}$ increases, electrons from bottom hadron
decays become important. The $p_T^{}$ at which electrons from
charmed hadron decays and from bottom hadron decays intercepts
depends on the model for heavy quark energy loss in the quark-gluon
plasma. Since heavy quark energy loss is suppressed as its mass
increases~\cite{DGVW05}, this momentum becomes smaller once heavy
quark energy loss is included. Therefore, semi-leptonic decays of
bottom hadrons should be considered at large electron $p_T^{}$
region, and this would further increase the electron $R_{AA}$.

\begin{figure}[t]\centering
\includegraphics[width=0.4\textwidth,angle=0,clip]{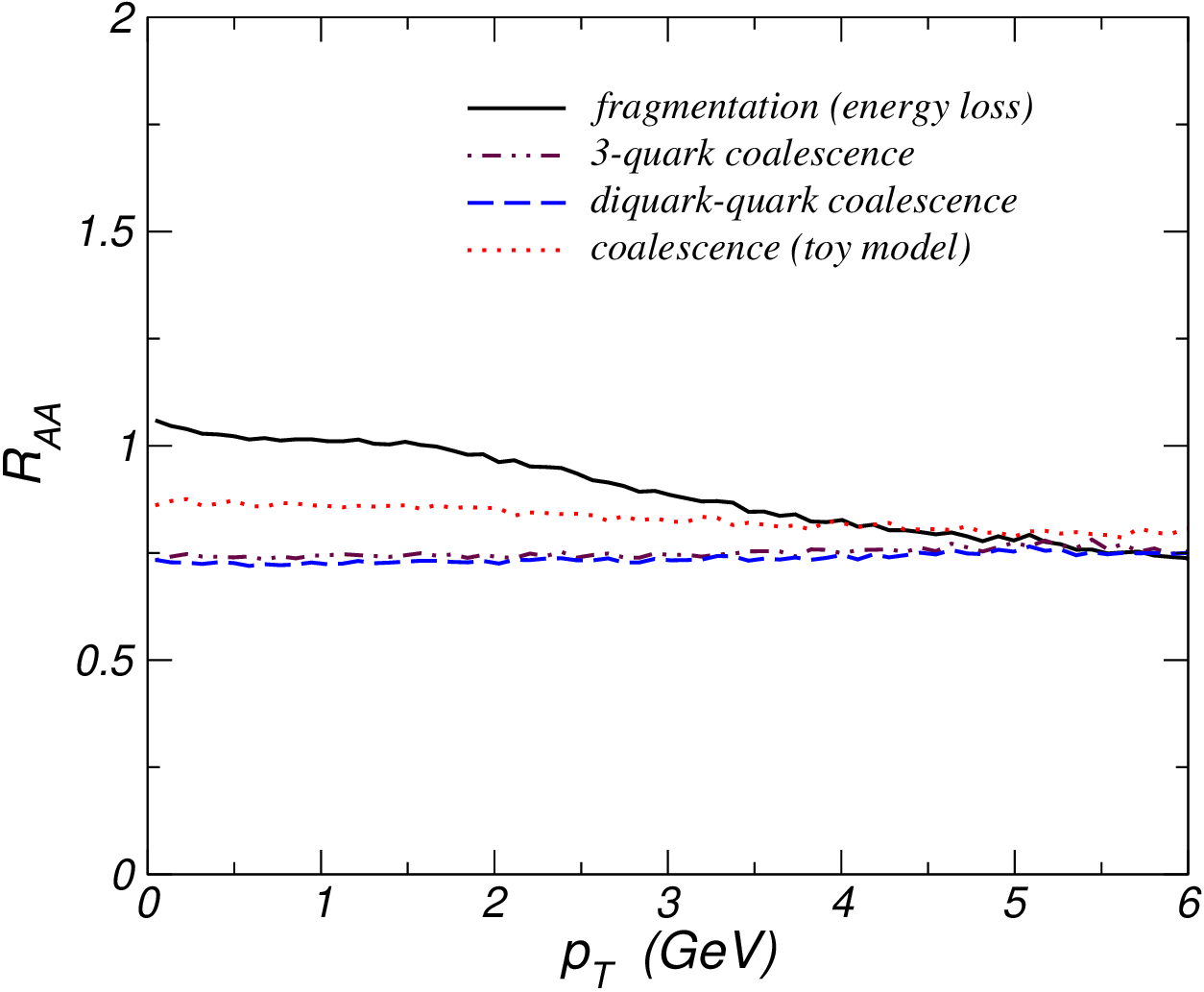}
\caption{\label{fig:RAAb}%
The electron $R_{AA}$ in the central collision of Au$+$Au at
$\sqrt{s_{NN}^{}} = 200$~GeV from bottom hadrons.
Notations are the same as in Fig.~\ref{fig:RAA}.
}
\end{figure}

The decays of bottom hadrons into electrons are not well-known.
There are two processes through which electrons are produced from
bottom hadron decays. One is the direct production of electrons from
bottom hadron decays, such as $B \to e \nu_e X_c$, where $X_c$
denotes any charm meson. The produced $X_c$ can again decay into
electrons. Therefore, bottom hadrons produce primary and secondary
electrons. In addition, for vector mesons, $B^*$ and $B_s^*$, they
first decay into $B\gamma$ and $B_s\gamma$, respectively, and
electrons are then produced from the decay of resulting pseudoscalar
mesons. Currently available experimental data show that the
branching ratios of bottom hadrons decaying into electrons are
nearly independent of the light quark content of bottom
hadrons~\cite{PDG08},
\begin{eqnarray}
\mbox{BR}(B^\pm \to e^+ \nu_e X_c) &=& 10.8 \pm 0.4\,\%,
\nonumber \\
\mbox{BR}(B^0 \to e^+ \nu_e X_c) &=& 10.1 \pm 0.4\,\%,
\nonumber \\
\mbox{BR}(B_s \to \ell^+ \nu_e D_s + \mbox{ anything}) &=& 7.9 \pm 2.4\,\%,
\nonumber \\
\mbox{BR}(\Lambda_b \to \ell^- \bar\nu_e \Lambda_c^+ + \mbox{
anything}) &=& 9.9 \pm 2.6\,\%.
\label{eq:bottom}
\end{eqnarray}
The paucity of experimental data for the branching ratio of each
channel in semi-leptonic decays of bottom hadrons makes it difficult
to compute the electron spectrum from bottom hadron decays.
Nevertheless, one can estimate the effect of $\Lambda_b/\bar{B}^0$
enhancement to the electron $R_{AA}$ in a qualitative way by using
the formula given in Eq.~(\ref{raa}). In Ref.~\cite{OKLY09}, we have
observed a large enhancement of the $\Lambda_b/\bar{B}^0$ ratio in
the coalescence model.  However, since the branching ratio of $B \to
e$ and $\Lambda_b \to e$ are similar, enhancement of $\Lambda_b$
baryon produces more electrons, which compensates almost the reduced
number of electrons from $B$ decays. This can also be verified
through Eq.~(\ref{raa}). Namely, if we use the branching ratios
given in Eq.~(\ref{eq:bottom}), $R_{AA}$ changes from $1.0$ with $C
= 0$ to $0.95$ with $C \to \infty$. This shows that the effect of
the $\Lambda_b/\bar{B}^0$ enhancement on the electron $R_{AA}$ is
very small compared to that for charm hadrons. This is shown
explicitly in Fig.~\ref{fig:RAAb}. By considering the large
uncertainty in the branching ratio of $\Lambda_b$ decays, the
effects of $\Lambda_b/\bar{B}^0$ enhancement is small, in particular
at large $p_T^{}$ region, where the electron $R_{AA}$ is expected to
be governed by bottom hadron decays.

Although the multiplicities of multi-heavy-quark baryons are very
small, production of heavy baryons containing one heavy quark and
one or two strange quarks is not negligible. In the coalescence
model of Ref.~\cite{OKLY09}, we found that the sum of the
multiplicities of $\Xi_Q$, $\Xi_Q'$, and $\Omega_Q$ is as large as
$30 \sim 50$\% of the multiplicity of $\Lambda_Q$, where $Q$ stands
for $c$ or $b$. However, the branching ratios of these hadrons have
not been measured and the electron spectrum from the decays of these
baryons thus can not be estimated.

In summary, we have studied the transverse momentum spectrum and
the nuclear modification factor $R_{AA}$ of non-photonic
electrons from heavy hadron decays in relativistic heavy ion
collisions. Contrary to the models of Refs.~\cite{SD05,MGC07}, which
assume an enhanced $\Lambda_c/D^0$ ratio at large
$p_T^{}$ region, the large enhancement of $\Lambda_c/D^0$ ratio
predicted in the coalescence model~\cite{OKLY09} occurs mainly in
the low $p_T^{}$ region and the heavy quark fragmentation remains
dominant at large $p_T^{}$ region. As a result, no additional
suppression to the electron $R_{AA}$ due to enhanced
$\Lambda_c/D^0$ ratio is obtained at large $p_T^{}$ region. We have
also estimated the role of $\Lambda_b/\bar{B}^0$ enhancement in the
electron $R_{AA}$ and found that the enhancement of $\Lambda_b$
baryon production does not affect the electron $R_{AA}$ produced by
bottom hadron decays because of similar branching ratios of bottom
meson and $\Lambda_b$ decays into electrons.

\acknowledgments

This work was supported by the US National Science Foundation under
Grants No.\ PHY-0457265 and PHY-0758115 and by the Welch Foundation
under Grant No.\ A-1358.

\end{document}